\documentclass[conference]{IEEEtran}
\usepackage{cite}
\usepackage{amsmath,amssymb,amsfonts}
\usepackage{algorithmic}
\usepackage{graphicx}
\usepackage{textcomp}
\usepackage{xcolor}

\usepackage{tikz}

\usepackage{tikz,ifthen,calc,graphicx}
\usepackage{pgfplots}

\usepackage{bm}
\usepackage{url}
\usepackage[font=small,labelfont=bf]{caption}
\usepackage[font=small,labelfont=bf]{subcaption}

\usepackage{lipsum}
\usepackage{makeidx}
\usepackage{enumerate}
\usepackage{color}
\usepackage{cite}
\usepackage{amsmath,amsthm}   
\usepackage{amssymb}
\usepackage{nomencl}
\usepackage{multirow}
\usepackage{epstopdf}
\usepackage{multicol}

\usepackage{calc}
\usepackage{here}

\usepackage{times}
\usepackage{dsfont}
\usepackage{epic,eepic}
\usepackage{rawfonts}
\usepackage{latexsym}
\usepackage{amsfonts}

\allowdisplaybreaks
\begin{document}
\title{Rate Region of MIMO RIS-assisted Broadcast Channels with Rate Splitting and Improper Signaling}

\author{%
   \IEEEauthorblockN{Mohammad Soleymani$^*$, Ignacio Santamaria$^\dag$, and Eduard Jorswieck$^\ddag$}
   \IEEEauthorblockA{*Signal \& System Theory Group, Universit\"at Paderborn, Germany \\
                     $^\dag$Dept. Communications Engineering, Universidad de Cantabria, Spain\\
$^\ddag$ Institute for Communications Technology, Technische Universit\"at Braunschweig, Germany\\
                     Email: \small{\protect\url{mohammad.soleymani}@sst.upb.de}\small{\protect\url{i.santamaria@unican.es}}, \small{\protect\url{jorswieck@ifn.ing.tu-bs.de}}
}
} 

\maketitle

\begin{abstract}
In this paper, we study the achievable rate region of 1-layer rate splitting (RS) in the presence of hardware impairment (HWI) and improper Gaussian signaling (IGS) for a single-cell reconfigurable intelligent surface (RIS) assisted broadcast channel (BC). We assume that the transceivers may suffer from an imbalance in in-band and quadrature signals, which is known as I/Q imbalance (IQI). The received signal and noise can be improper when there exists IQI. Therefore, we employ IGS to compensate for IQI  as well as to manage interference. Our results show that RS and RIS can significantly enlarge the rate region, where the role of RS is  to manage interference while RIS mainly improves the coverage.  
\end{abstract} 

\section{Introduction}
The sixth generation (6G) of wireless communication systems should be much more spectral and energy efficient than the existing communication systems \cite{chafii2022ten}.
This goal may not  be achieved without employing some emerging 
 technologies such as reconfigurable intelligent surfaces (RISs) and rate splitting (RS), which have been shown to be able to improve the spectral and energy efficiency of various wireless communication systems \cite{wu2021intelligent, di2020smart, mao2022rate, clerckx2022primer}. 

Interference has been always among the main restrictions of modern wireless communication systems, and interference-management techniques are thus  expected to continue playing a key role in such systems \cite{andrews2014will}.
Even though there are many studies on the performance of RIS, its role should be further investigated in overloaded interference-limited systems, in which the number of users is larger than the number of spatial, temporal, or frequency resources.
RIS can modulate channels, canceling interference links or improving the strength of desired links. In other words, RIS can be potentially employed to manage and/or neutralize interference in some scenarios such as multiple-user interference channels. Thus, the following question may arise: 
Are other advanced interference management techniques, such as rate-splitting or improper Gaussian signaling (IGS), still necessary in RIS-assisted systems? We answer this question in positive in this paper. 

Rate splitting (RS) is a powerful technology to highly improve the spectral and energy efficiency of various interference-limited systems \cite{mao2022rate, clerckx2022primer}. There are different RS schemes such as 1-layer RS, hybrid RS and generalized RS. The generalized RS scheme is the most complete RS scheme and includes many other technologies/techniques such as spatial division multiple access (SDMA), non-orthogonal multiple access (NOMA), orthogonal multiple access (OMA) and treating interference as noise (TIN) \cite{mao2018rate}. 
Implementing generalized RS has high complexities when the number of users grows. 
An alternative to the generalized RS is 1-layer RS, which is a very practical scheme with much lower complexities. 1-layer RS is very efficient and is able to improve the performance of different interference-limited systems \cite{clerckx2019rate, mao2020beyond, flores2020linear}. 

Another powerful interference-management technique is improper Gaussian signaling (IGS), which can improve the system performance when the receivers apply TIN or partial successive cancellation (SIC) \cite{javed2018improper, soleymani2020improper,  soleymani2019improper,  soleymani2019ergodic, nasir2020signal, yu2021maximizing}. Moreover, IGS has been shown to increase the degrees-of-freedom of the $3$-user single-input, single-output (SISO) interference channel (IC)  through interference alignment \cite{cadambe2010interference}.  

Interference is not the only performance limiting factor in wireless communication systems.
Another limitation arises from non-idealities in transceivers. A source of imperfections in transceivers is I/Q imbalance (IQI), which is modeled as a widely linear transformation of the input signal \cite{javed2019multiple, soleymani2020improper, boulogeorgos2016energy}. Hardware impairments (HWI) can highly affect the system performance especially when such imperfections are not taken into account in the system design. 

In this paper, we investigate the role of RIS, RS and IGS in multiple-input, multiple-output (MIMO) broadcast channels (BCs) with IQI. We show that although RIS can enlarge the rate region, RS and IGS are still needed to manage interference and compensate for IQI. Indeed, the role of RIS in this scenario is mainly to improve the coverage, while RS is responsible for handling interference. It is known that 1-layer RS with proper Gaussian signaling (PGS) is the optimal scheme in the two-user single-cell BC with perfect devices. However, it is not the case in the presence of IQI.  Our results show that IQI shrinks the achievable rate region, and IGS with TIN  may outperform RS with PGS in some operational points/regimes. In this case, the 1-layer RS with IGS outperforms the other considered schemes. 

This paper is organized as follows. Section \ref{secII} presents the system model. Section \ref{sec-iii} proposes a suboptimal scheme to obtain the achievable rate region. 
Section \ref{sec-iv} presents some numerical results, and Section \ref{v} summarizes the main findings of the paper.

\section{System Model}\label{secII}

\subsection{Network Scenario}\label{secII-A}
\begin{figure}[t!]
    \centering
\includegraphics[width=.35\textwidth]{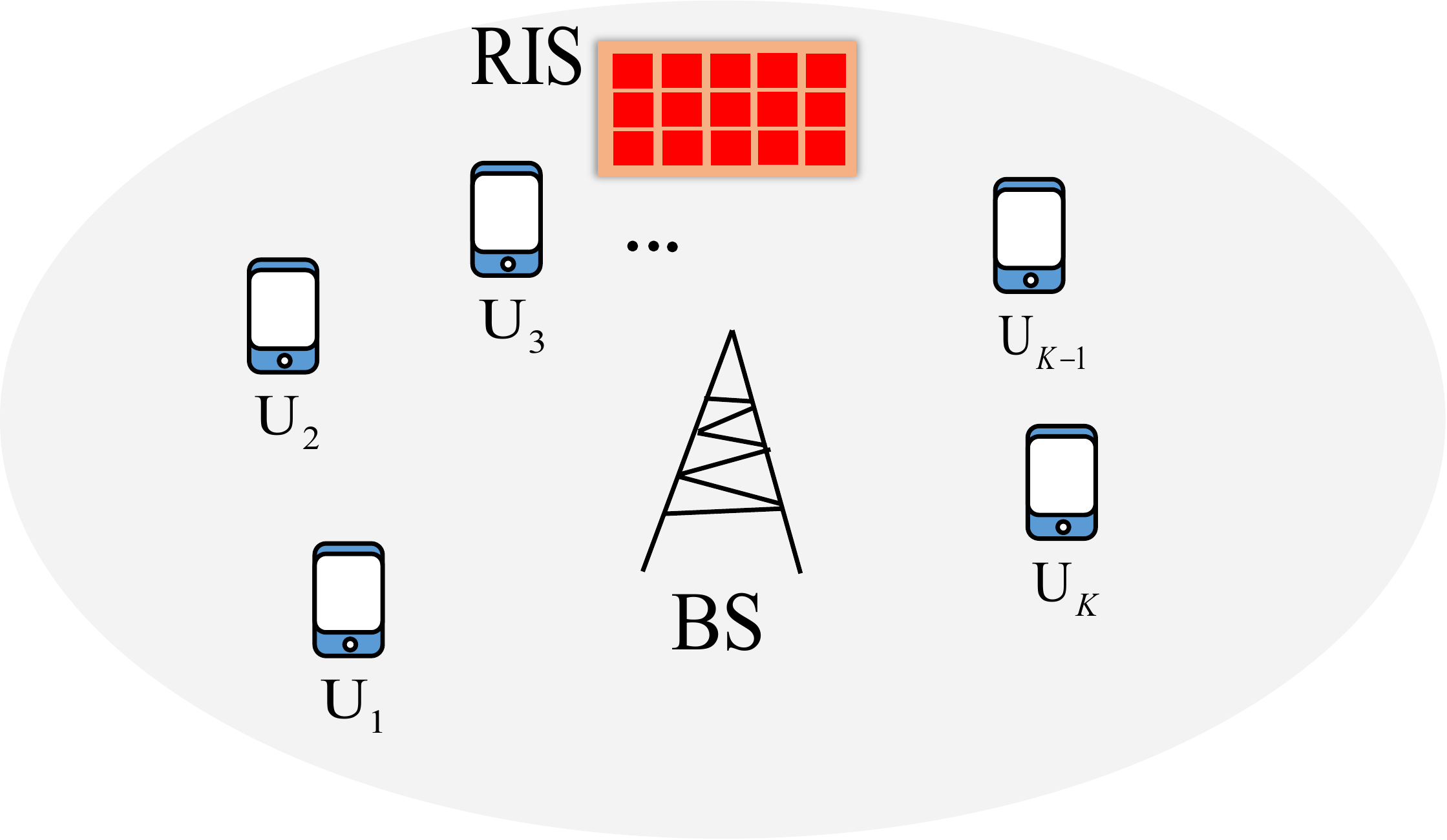}
     \caption{A broadcast channel with RIS.}
	\label{Fig-sys-model}
\end{figure}

We consider a single-cell 
RIS-assisted system with IQI at transceivers, as shown in Fig. \ref{Fig-sys-model}. We assume that there is a BS with $N_{BS}$ transmit antennas, serving $K$ users with $N_u$ receive antennas  each. Additionally, there is a RIS with $N_{RIS}$ components to assist the BS. The users and the BS may suffer from IQI according to the model in  \cite{javed2019multiple, soleymani2020improper}.   
For the sake of notational simplicity, we consider a symmetric scenario in which all the users have the same number of antennas as well as the same IQI parameters, although the model can be easily extended to asymmetric scenarios.
\subsection{Channel model}
We employ the channel model in \cite{pan2020multicell} for MIMO RIS-assisted systems. In this  paper, we briefly present the channel model and refer the readers to \cite{pan2020multicell, soleymani2022rate} for more detailed discussions on the fading models of MIMO RIS-assisted systems. 
The channel matrix between the BS and user $k$ is \cite[Eq. (14)]{soleymani2022improper}  
\begin{equation}\label{ch-equ}
\mathbf{H}_{k}\left(\bm{\Theta}\right)=
\underbrace{\mathbf{G}_{k}\bm{\Theta}\mathbf{G}}_{\text{Link through RIS}}
+
\underbrace{\mathbf{F}_{k}}_{\text{Direct link}}
\in\mathbb{C}^{N_u\times N_{BS}},
\end{equation}
where $\mathbf{F}_{k}$ is the channel matrix between the BS and user $k$,  
$\mathbf{G}_{k}$ is the channel matrix between the RIS and user $k$, 
$\mathbf{G}$ is the channel matrix between the BS and the RIS, and the matrix $\bm{\Theta}$ is 
\begin{equation}
\bm{\Theta}=\text{diag}\left(\theta_{1}, \theta_{2},\cdots,\theta_{{N_{RIS}}}\right),
\end{equation}
where $\theta_i$s for all $i$ are RIS components. 
In this paper, amplitudes of the RIS components are assumed to be fixed to $1$, while the phases can take any value between $0$ and $2\pi$. In other words, the constraint set for the RIS components is
\begin{equation}
\mathcal{T}=\left\{\theta_{i}:|\theta_{i}|= 1 \,\,\,\forall i\right\}.
\end{equation}
We refer the reader to \cite{soleymani2022rate} for a description of other common constraints on the amplitudes and phases of the RIS elements. 

\subsection{1-layer rate splitting scheme}
We consider 1-layer RS scheme in which the BS broadcasts a common message for all users and $K$ private messages (one for each user). Thus, the BS is intended to transmit  
\begin{equation}
\mathbf{x}=\mathbf{x}_{c}+\sum_{k=1}^K\mathbf{x}_{k}, 
\end{equation}
where $\mathbf{x}_{c}$ is the common message, and $\mathbf{x}_{k}$ is the private message of the BS intended for user $k$.  Since the BS may suffer from IQI, the actual transmit signal of the BS is a widely linear transformation of $\mathbf{x}$ as \cite{javed2019multiple, soleymani2020improper}
\begin{equation}
\mathbf{x}_t=\mathbf{V}_{1}\mathbf{x}+\mathbf{V}_{2}\mathbf{x}^*, 
\end{equation}
where the constant matrices $\mathbf{V}_{1}$ and $\mathbf{V}_{2}$ are, respectively, defined in \cite[Eq. (7)]{soleymani2020improper} and \cite[Eq. (8)]{soleymani2020improper}. 
The receive signal at the receiver of user $k$ is
\begin{equation}
\mathbf{y}_k=\mathbf{H}_k\mathbf{x}_t+\mathbf{n}_{k}, 
\end{equation}
where $\mathbf{n}_{k}$ is a zero-mean proper white additive Gaussian noise with variance $\sigma^2$ at receiver $k$, and $\mathbf{H}_k$ is the effective channel between the BS and user $k$, given by \eqref{ch-equ}. Note the effective channel is a function of $\bm{\Theta}$; however, we drop the dependency to simplify the representation of equations. 
The receiver of user $k$ may suffer from  IQI, which means the final output of the received signal is a widely linear transformation of $\mathbf{y}_k$ as
\begin{multline}\label{eq-1}
\mathbf{y}_k=\mathbf{\Gamma}_{1}\left[\mathbf{H}_k\left(\mathbf{V}_{1}\mathbf{x}_k+\mathbf{V}_{2}\mathbf{x}^*_k\right)+\mathbf{n}_k\right]
\\
+\mathbf{\Gamma}_{2}\left[\mathbf{H}_k\left(\mathbf{V}_{1}\mathbf{x}_k+\mathbf{V}_{2}\mathbf{x}^*_k\right)+\mathbf{n}_k\right]^*,
\end{multline}
where the constant matrices $\mathbf{\Gamma}_{1}$ and $\mathbf{\Gamma}_{2}$ are given by \cite[Eq. (12)]{soleymani2020improper} and \cite[Eq. (13)]{soleymani2020improper}, respectively. 
Note that the effective noise at user $k$ can be improper due to IQI, meaning that the real and imaginary parts of the effective noise can be correlated and/or have unequal powers \cite{schreier2010statistical}.
To compensate for IQI, we assume that the common and private messages, $\mathbf{x}_{c}$ and $\mathbf{x}_{k}$, can be improper Gaussian signals. Additionally, the signals $\mathbf{x}_{c}$ and $\mathbf{x}_{k}$ are zero-mean and independent. 
Note that IGS can also help to reduce or manage interference, as indicated before, so its role is not only to compensate for IQI.

A way to model IGS is through the real-decomposition method. We can rewrite \eqref{eq-1} by employing the real-decomposition method as
\begin{align}
\nonumber
\underline{\mathbf{y}}_{k}&=\underline{\mathbf{H}}_{k}
\underline{\mathbf{x}}_{i}+\underline{\mathbf{n}}_{lk}
\\&
=
\underbrace{\underline{\mathbf{H}}_{k}
\underline{\mathbf{x}}_{c}}_{\text{Common  M.}}+
\underbrace{\underline{\mathbf{H}}_{k}
\underline{\mathbf{x}}_{k}}_{\text{Private M.}}
+
\underbrace{\underline{\mathbf{H}}_{k}
\sum_{j=1,j\neq k}^{K}\underline{\mathbf{x}}_{j}}_{\text{Interference}}
+
\underbrace{\underline{\mathbf{n}}_{k}}_{\text{Noise}},
\end{align}
where 
$\underline{\mathbf{H}}_{k}$ is the equivalent channel given by \cite[Eq. (11)]{soleymani2022improper}, 
$\underline{\mathbf{y}}=\left[ \begin{array}{cc}
\mathfrak{R}\{\mathbf{y}\}^T & \mathfrak{I}\{\mathbf{y}\}^T \end{array} \right]^T$,  and
$\underline{\mathbf{x}}=\left[ \begin{array}{cc}
\mathfrak{R}\{\mathbf{x}\}^T & \mathfrak{I}\{\mathbf{x}\}^T \end{array} \right]^T$
 are, respectively, the real decomposition of $\mathbf{y}$ and $\mathbf{x}$. 
Additionally, $\underline{\mathbf{n}}_k$ represents the effective improper noise and is given by 
\begin{equation}
\underline{\mathbf{n}}_k=\left[ \begin{array}{cc}
\mathfrak{R}\{\mathbf{\Gamma}_{1}\mathbf{n}+\mathbf{\Gamma}_{2}\mathbf{n}^*\}^T & \mathfrak{I}\{\mathbf{\Gamma}_{1}\mathbf{n}+\mathbf{\Gamma}_{2}\mathbf{n}^*\}^T \end{array} \right]^T.
\end{equation} 
We represent the covariance matrix of the noise by $\mathbb{E}\{\underline{\mathbf{n}}\,\underline{\mathbf{n}}^T\}=\underline{\mathbf{C}}_{n}$. 
Moreover, the transmit covariance matrix of  $\underline{\mathbf{x}}$, $\underline{\mathbf{x}}_{c}$, and $\underline{\mathbf{x}}_{k}$ are denoted as ${\bf P}$, ${\bf P}_{c}$, and ${\bf P}_{k}$, respectively, where $\mathbf{P}=\mathbf{P}_{c}+\sum_{k}\mathbf{P}_{k}$. 
The constraint set of the transmit covariance matrices is given by \cite[Eq. (3)]{soleymani2022rate} for IGS (denoted by $\mathcal{P}_I$) and by \cite[Eq. (4)]{soleymani2022rate} for PGS schemes (denoted by $\mathcal{P}_P$). 
Since RS can be applied to both PGS and IGS cases, we represent the constraint set of the transmit covariance matrices as $\mathcal{P}$. 
Note that the set  $\mathcal{P}$ is convex. Moreover, note that  PGS is a special case of IGS, thus, an optimal IGS scheme never performs worse than a PGS scheme.
We refer the reader to \cite[Sec. II.A]{soleymani2022improper}, \cite[Appendix A]{soleymani2022rate} and \cite{schreier2010statistical} for further details on modeling IQI and/or impropriety.

 \subsection{Rate expressions}
Users firstly decode the common message and cancel it from the received signal. Thus, the maximum decoding rate for the common message at user $k$ is
\cite[Eqs. (2)-(3)]{mishra2021rate}
\begin{align}
\nonumber
\bar{r}_{ck}&=\frac{1}{2}\log_2\left|{\bf I}+
\left(
\underline{\mathbf{C}}_{n}+
\sum_{\forall i}\underline{{\bf H}}_{k}
{\bf P}_{i}\underline{{\bf H}}_{k}^T\right)^{-1}
 \underline{{\bf H}}_{k}
{\bf P}_{c}\underline{{\bf H}}_{k}^T  
\right|
\\ &\nonumber
=
\underbrace{
\frac{1}{2}\log_2\left|
\underline{\mathbf{C}}_{n}+
\underline{{\bf H}}_{k}
{\bf P}\underline{{\bf H}}_{k}^T
\right|
}_{\bar{r}_{ck,1}
}
-
\underbrace{
\frac{1}{2}\log_2\left|
\underline{\mathbf{C}}_{n}+
\sum_{\forall i}\underline{{\bf H}}_{k}
{\bf P}_{i}\underline{{\bf H}}_{k}^T
\right|}
_{\bar{r}_{ck,2}
}.
\label{eq=29=}
\end{align}
 The common message must be transmitted at a rate that is decodable for all users. Hence, the maximum rate for transmitting the common message is   
\begin{equation}\label{fis=}
r_c(\{\mathbf{P}\},\bm{\Theta})= \min_{
k}\left\{\bar{r}_{ck}(\{\mathbf{P}\},\bm{\Theta})\right\}.
\end{equation}
After decoding and canceling the common message, each user decodes its own private message. Therefore, the maximum decoding rate for the private message at user $k$ is 
\begin{align}\label{eq-28}
r_{pk}\!&=\!
\frac{1}{2}\log_2\left|{\bf I}+
\!\left(
\!\underline{\mathbf{C}}_{n}+
\!\sum_{\forall i\neq k}\underline{{\bf H}}_{k}
{\bf P}_{i}\underline{{\bf H}}_{k}^T
\!\right)^{-1}
\!\! \underline{{\bf H}}_{k}
{\bf P}_{k}\underline{{\bf H}}_{k}^T  
\right|
\\ &
=\!
\underbrace{
\frac{1}{2}\!\log_2\!\left|
\underline{\mathbf{C}}_{n}\!\!+\!\!
\sum_{\forall i}\!\underline{{\bf H}}_{k}
{\bf P}_{i}\underline{{\bf H}}_{k}^T
\right|
}_{r_{pk,1}
}
\!-\!
\underbrace{
\frac{1}{2}\!\log_2\!\left|
\underline{\mathbf{C}}_{n}\!\!+\!\!\!
\sum_{\forall i\neq k}\!\underline{{\bf H}}_{k}
{\bf P}_{i}\underline{{\bf H}}_{k}^T
\right|
}_{r_{pk,2}
}\!.
\end{align}
Finally, the rate of user $k$ is the summation of the decoding rate of its private message and 
its dedicated rate from the common message, i.e., \cite{soleymani2022rate}
\begin{equation}
r_k(\{\mathbf{P}\},\bm{\Theta})=r_{pk}(\{\mathbf{P}\},\bm{\Theta})+r_{ck},
\end{equation}  
where $r_{ck}\geq 0$ and 
$\sum_k r_{ck}\leq r_c$. Note that the rates $r_k$, $r_{pk}$, $\bar{r}_{ck}$ and $r_c$  are functions of $\{\mathbf{P}\}$ and $\bm{\Theta}$, while $\mathbf{r}_c=\left\{r_{c1},r_{c2},\cdots,r_{cK}\right\}$ is a design parameter. Due to notational simplicity, we, hereafter, drop the dependency of the rates to $\{\mathbf{P}\}$ and $\bm{\Theta}$ in representing $r_k$, $r_{pk}$, $\bar{r}_{ck}$ and $r_c$.
\subsection{Problem Statement}
Employing the rate profile technique, the achievable rate region can be obtained by solving \cite{soleymani2020rate}
\begin{subequations}\label{rr}
\begin{align}
\max_{\bm{\Theta}\in\mathcal{T},\{\mathbf{P}\}\in\mathcal{P},\mathbf{r}_c}\,\, &r
\\
\label{12b}
\text{s.t.}\,\,&r_{k}=r_{pk}+r_{ck} \geq \alpha_kr&\forall k,\\
\label{12c}
&\sum_{\forall k}r_{ck}\leq r_c, \hspace{.3cm}r_{ck}\geq 0,&\forall k,
\end{align}
\end{subequations}
and varying the weights such that $\sum_{\forall k}\alpha_k=1$ with $\alpha_k\geq 0$ for all $k$.

\section{Proposed algorithm}\label{sec-iii}
We employ an approach based on the optimization framework proposed in \cite{soleymani2022rate} to solve \eqref{rr}. That is, we first employ an alternating optimization (AO) approach to sequentially optimize over the transmit covariance matrices and RIS components. Indeed, we first fix the RIS components to $\bm{\Theta}^{(t-1)}$ and update the transmit covariance matrices as $\{\mathbf{P}^{(t)}\}$. Then, we alternate and optimize over the reflecting coefficients for fixed transmit covariance matrices. Unfortunately, even after fixing either the RIS or the covariance matrices, the resulting optimization problems are still non-convex. In the following subsections, we proposed iterative algorithms to find a suboptimal solution. 

\subsection{Optimizing transmit covariance matrices}  
In this subsection, we update the transmit covariance matrices for a fixed $\bm{\Theta}^{(t-1)}$ by solving
\begin{align}\label{13}
\max_{\{\mathbf{P}\}\in\mathcal{P},\mathbf{r}_c}\,\, &r&
\text{s.t.}\,\,&\,\,\,
\eqref{12b},\eqref{12c},
\end{align}
which is a non-convex problem. 
Note that the constraints in \eqref{13} are linear in  $\mathbf{r}_c$, however, the rates are not concave in $\{\mathbf{P}\}$, which makes the problem non-convex.
Indeed, the rate $r_{pk}$ (or $r_{ck}$) can be written as a difference of two concave functions $r_{pk,1}$ and $r_{pk,2}$ (or $r_{ck,1}$ and $r_{ck,2}$).
Thus, to solve \eqref{13}, we can employ difference of convex programming (DCP), which falls into majorization minimization (MM). That is, we approximate the rates by a suitable concave lower bound. To this end, we keep  $r_{pk,1}$ (or $r_{ck,1}$) unchanged and employ the first-order Taylor expansion to approximate $r_{pk,2}$ (or $r_{ck,2}$) by an affine (linear) function as
\begin{multline}
\label{l-r-lk-p}
 r_{pk}\geq \tilde{r}_{pk}= 
r_{pk,1}
-r_{pk,2}\left(\{\mathbf{P}^{(t-1)}\}\right)
\\
-\!\!\sum_{\forall j\neq k}\!\!\!
\text{Tr}\!\!\left(\!\!
\frac{
\underline{\mathbf{H}}_{k}^T
(\underline{\mathbf{C}}_{n}\!\!+\!\!
\sum_{\forall i\neq k}\!\underline{{\bf H}}_{k}
{\bf P}_{i}^{(t-1)}\underline{{\bf H}}_{k}^T)^{-1}
\underline{\mathbf{H}}_{k}
}
{2\ln 2}\!\!
\left(\!\!\mathbf{P}_{j}\!-\!\mathbf{P}_{j}^{(t-1)}\!\right)
\!\!\!\right)\!\!.
\end{multline}
Similarly, a concave lower bound for $\bar{r}_{ck}$ is 
\begin{multline}
\label{l-r-ck}
{r}_{ck}\geq \tilde{r}_{ck}= 
\bar{r}_{ck,1}
-\bar{r}_{ck,2}\left(\{\mathbf{P}^{(t-1)}\}\right)
\\
-\!\!\sum_{\forall j}\!\!
\text{Tr}\!\!\left(\!\!
\frac{
\underline{\mathbf{H}}_{k}^T
(\underline{\mathbf{C}}_{n}\!\!+\!\!
\sum_{\forall i}\!\underline{{\bf H}}_{k}
{\bf P}_{i}^{(t-1)}\underline{{\bf H}}_{k}^T)^{-1}
\underline{\mathbf{H}}_{k}
}
{2\ln 2}\!\!
\left(\!\!\mathbf{P}_{j}\!-\!\mathbf{P}_{j}^{(t-1)}\!\right)
\!\!\!\right)\!\!.
\end{multline}
 We refer the reader to \cite[Corollary 1]{soleymani2022rate} for the proof. 
Substituting the concave lower bounds in \eqref{13} results in the following convex optimization problem
\begin{subequations}\label{rr-s}
\begin{align}
\max_{\{\mathbf{P}\}\in\mathcal{P},\mathbf{r}_c}\,\, &r
\\
\text{s.t.}\,\,&\tilde{r}_{k}=\tilde{r}_{pk}+r_{kc} \geq \alpha_kr&\forall k,\\
&\sum_{\forall k}r_{ck}\leq \tilde{r}_c=\min_k\left\{\tilde{r}_{ck}\right\}, \hspace{.3cm}r_{ck}\geq 0,&\forall k.
\end{align}
\end{subequations}
This problem can be solved by existing numerical tools, which yields $\{\mathbf{P}^{(t)}\}$.
\subsection{Optimizing RIS components}
Now we update the RIS components $\bm{\Theta}$ for fixed transmit covariance matrices $\{\mathbf{P}^{(t)}\}$ by solving 
\begin{align}\label{14}
\max_{\bm{\Theta}\in\mathcal{T},\mathbf{r}_c}\,\, &r&
\text{s.t.}\,\,&\,\,\,
\eqref{12b},\eqref{12c},
\end{align}
This problem is non-convex since the constraint set for the RIS components is not a convex set, and additionally, the rates are not concave in $\bm{\Theta}$. Thus, to solve \eqref{14}, we first find suitable concave lower bounds for the rates and then, convexify the  constraint set $\mathcal{T}$. To this end, we employ the concave lower bound given by \cite[Lemma 4]{soleymani2022rate}, which results in the 
following concave lower bound for $r_{pk}$ 
\begin{multline} 
r_{pk}\geq
 \hat{r}_{pk}=r_{pk}\left(\bm{\Theta}^{(t-1)}\right)-
\frac{1}{2\ln 2}\text{Tr}\left(
\bar{\mathbf{V}}_{k}\bar{\mathbf{V}}_{k}^T\bar{\mathbf{Y}}_{k}^{-1}
\right)
\\-
\frac{1}{2\ln 2}
\text{Tr}\left(
[\bar{\mathbf{Y}}^{-1}_{k}\!-(\bar{\mathbf{V}}_{k}\bar{\mathbf{V}}^T_{k}\! +\! \bar{\mathbf{Y}}_{k})^{-1}]^T
[\mathbf{V}_{k}\mathbf{V}^T_{k}\!+\!\mathbf{Y}_{k}]
\right)
\\
+
\frac{1}{\ln 2}
\text{Tr}\left(
\bar{\mathbf{V}}_{k}^T\bar{\mathbf{Y}}_{k}^{-1}\mathbf{V}_{k}
\right),
\end{multline}
where 
 $\mathbf{V}_{k}=\underline{\mathbf{H}}_{k}\left(\bm{\Theta}\right)\mathbf{P}_{k}^{(t)^{1/2}}$, $\bar{\mathbf{V}}_{k}=\underline{\mathbf{H}}_{k}\left(\bm{\Theta}^{(t-1)}\right)\mathbf{P}_{k}^{(t)^{1/2}}$ and
\\
\begin{align}
\mathbf{Y}_{k}&=\underline{\mathbf{C}}_{n}+
\!\sum_{\forall i\neq k}\underline{{\bf H}}_{k}\left(\bm{\Theta}\right)
{\bf P}_{i}^{(t)}\underline{{\bf H}}_{k}^T\left(\bm{\Theta}\right)
\\
\bar{\mathbf{Y}}_{k}&=\underline{\mathbf{C}}_{n}+
\!\sum_{\forall i\neq k}\underline{{\bf H}}_{k}\left(\bm{\Theta}^{(t-1)}\right)
{\bf P}_{i}^{(t)}\underline{{\bf H}}_{k}^T\left(\bm{\Theta}^{(t-1)}\right).
\end{align}
Similarly, a concave lower bound for $\bar{r}_{ck}$ can be found as
\begin{multline} 
\bar{r}_{ck}\geq
 \hat{r}_{ck}=\bar{r}_{ck}\left(\bm{\Theta}^{(t-1)}\right)-
\frac{1}{2\ln 2}\text{Tr}\left(
\bar{\mathbf{V}}_{ck}\bar{\mathbf{V}}_{ck}^T\bar{\mathbf{Y}}_{ck}^{-1}
\right)
\\-\!
\frac{1}{2\ln 2}\!
\text{Tr}\left(
[\bar{\mathbf{Y}}^{-1}_{ck}\!-(\bar{\mathbf{V}}_{ck}\bar{\mathbf{V}}^T_{ck}\! +\! \bar{\mathbf{Y}}_{ck})^{-1}]^T
[\mathbf{V}_{ck}\mathbf{V}^T_{ck}\!+\!\mathbf{Y}_{ck}]
\right)
\\
+
\frac{1}{\ln 2}
\text{Tr}\left(
\bar{\mathbf{V}}_{ck}^T\bar{\mathbf{Y}}_{ck}^{-1}\mathbf{V}_{ck}
\right),
\end{multline}
where 
 $\mathbf{V}_{ck}=\underline{\mathbf{H}}_{k}\left(\bm{\Theta}\right)\mathbf{P}_{c}^{(t)^{1/2}}$, $\bar{\mathbf{V}}_{ck}=\underline{\mathbf{H}}_{k}\left(\bm{\Theta}^{(t-1)}\right)\mathbf{P}_{c}^{(t)^{1/2}}$ and
\\
\begin{align}
\mathbf{Y}_{ck}&=\underline{\mathbf{C}}_{n}+
\!\sum_{\forall i}\underline{{\bf H}}_{k}\left(\bm{\Theta}\right)
{\bf P}_{i}^{(t)}\underline{{\bf H}}_{k}^T\left(\bm{\Theta}\right)
\\
\bar{\mathbf{Y}}_{ck}&=\underline{\mathbf{C}}_{n}+
\!\sum_{\forall i}\underline{{\bf H}}_{k}\left(\bm{\Theta}^{(t-1)}\right)
{\bf P}_{i}^{(t)}\underline{{\bf H}}_{k}^T\left(\bm{\Theta}^{(t-1)}\right).
\end{align}

Now we propose a suboptimal approach to convexify the constraint $|\theta_{n}|=1$ for all $n$. This constraint can be rewritten as 
$|\theta_{n}|^2\leq 1$, and 
$|\theta_{n}|^2\geq 1$, where the former is convex, while the latter is not. 
To convexify $|\theta_{n}|^2\geq 1$, we employ the first-order Taylor expansion since $|\theta_{n}|^2$ is a convex function, which results in
\begin{equation}\label{26}
|\theta_{n}|^2\geq|\theta_{n}^{(t-1)}|^2-2\mathfrak{R}\{\theta_{n}^{(t-1)^*}(\theta_{n}-\theta_{n}^{(t-1)})\}\geq 1.
\end{equation}
To converge faster, we relax the constraint in \eqref{26} as
\begin{equation}\label{+=+}
|\theta_{n}|^2\!\geq\!|\theta_{n}^{(t-1)}|^2\!-2\mathfrak{R}\{\theta_{n}^{(t-1)^*}(\theta_{n}-\theta_{n}^{(t-1)})\}\!\geq 1-\epsilon,
\end{equation}
where $\epsilon>0$. The constraint \eqref{+=+} is linear. Hence, the following surrogate optimization problem is convex
\begin{subequations}\label{rr-s2}
\begin{align}
\max_{\bm{\Theta},\mathbf{r}_c}\,\, &r
\\
\text{s.t.}\,\,&\hat{r}_{k}=\hat{r}_{pk}+r_{kc} \geq \alpha_kr&\forall k,\\
&\sum_{\forall k}r_{ck}\leq \hat{r}_c=\min_k\left\{\hat{r}_{ck}\right\}, \hspace{.3cm}r_{ck}\geq 0,&\forall k,\\
&|\theta_n|^2\leq 1,\hspace{.3cm}\text{and}\hspace{.3cm}\eqref{+=+},&\forall n.
\end{align}
\end{subequations}
 The solution of \eqref{rr-s2}, $\hat{\bm{\Theta}}$, may not necessarily be in $\mathcal{T}$ since we relaxed the constraint in \eqref{26}. To get a feasible point, we normalize $\hat{\bm{\Theta}}$ as 
$\theta_{n}^{\text{new}}=
{\hat{\theta}_{n}}/{|\hat{\theta}_{n}|}$,  for all $n$.
 Finally, we update ${\bm{\Theta}}$ based on the following rule
\begin{equation}\label{eq-42}
\bm{\Theta}^{(t)}=
\left\{
\begin{array}{lcl}
\hat{\bm{\Theta}}^{\text{new}}&\text{if}&
\min_{k}\left\{
\frac{
r_k\left(\left\{\mathbf{P}^{(t)}\right\},\hat{\bm{\Theta}}^{\text{new}}\right)
}
{\alpha_k}
\right\}\geq
\\
&&
\min_{k}\left\{
\frac{
r_k\left(\left\{\mathbf{P}^{(t)}\right\},\hat{\bm{\Theta}}^{(t-1)}\right)
}
{\alpha_k}
\right\}
\\
\{\bm{\Theta}^{(t-1)}\}&&\text{Otherwise}.
\end{array}
\right.
\end{equation}
This updating rule guarantees the convergence since the algorithm generates a sequence of non-decreasing minimum weighted rates.

\section{Numerical results}\label{sec-iv}
In this section, we provide some numerical examples to clarify the role of RS, RIS and IGS in single-cell BCs. We consider a line-of-sight (LoS) connection for the links reaching to or departing from the RIS, 
and a non-LoS (NLoS) link for the direct links between the users and the BS. It means that the small-scale fading of the links related to RIS is Rician, while that of direct links is Rayleigh. The large-scale path loss component of RIS links is $\alpha_{RIS}=3.2$. The other simulation parameters are chosen based on \cite{soleymani2022improper}. 
The considered schemes in the simulations are as follows: {\bf PT} (or {\bf IT}) denotes the PGS (or IGS) scheme with TIN but without RIS. {\bf PR} (or {\bf IR}) denotes the PGS-based (or IGS-based) RS scheme without RIS. {\bf PR$_I$R} (or {\bf IR$_I$R}) denotes the PGS-based (or IGS-based) RS scheme with RIS. Finally, {\bf TS} denotes the time-division-multiplexing-access (TDMA) with time sharing.   
\subsection{SISO systems}\label{sec-rr-4-a}
\begin{figure}[t!]
    \centering
    \begin{subfigure}[t]{0.24\textwidth}
        \centering
        \includegraphics[width=.96\textwidth]{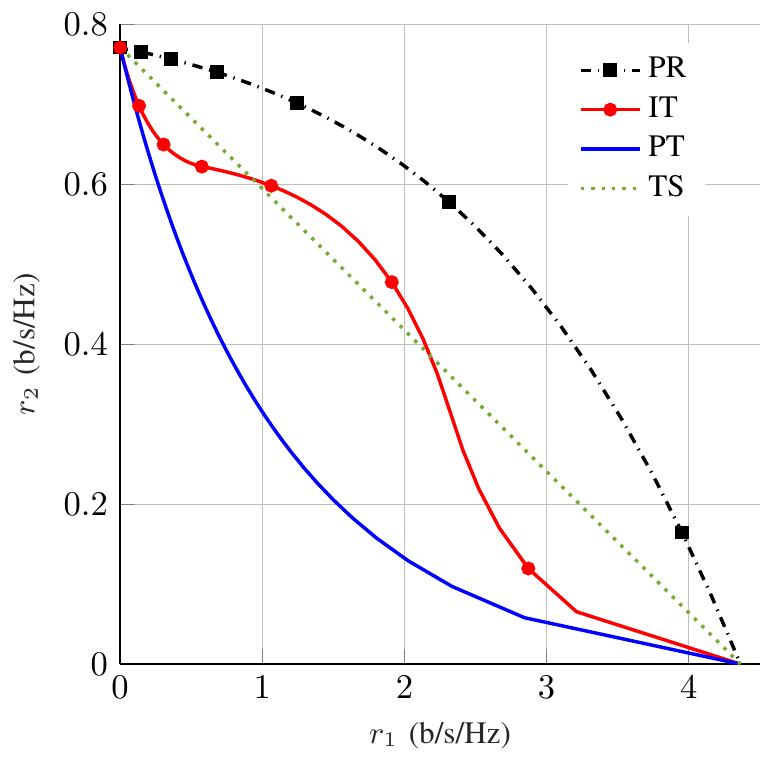}
        \caption{With perfect devices.}
    \end{subfigure}%
    ~ 
    \begin{subfigure}[t]{0.24\textwidth}
        \centering
\includegraphics[width=.96\textwidth]{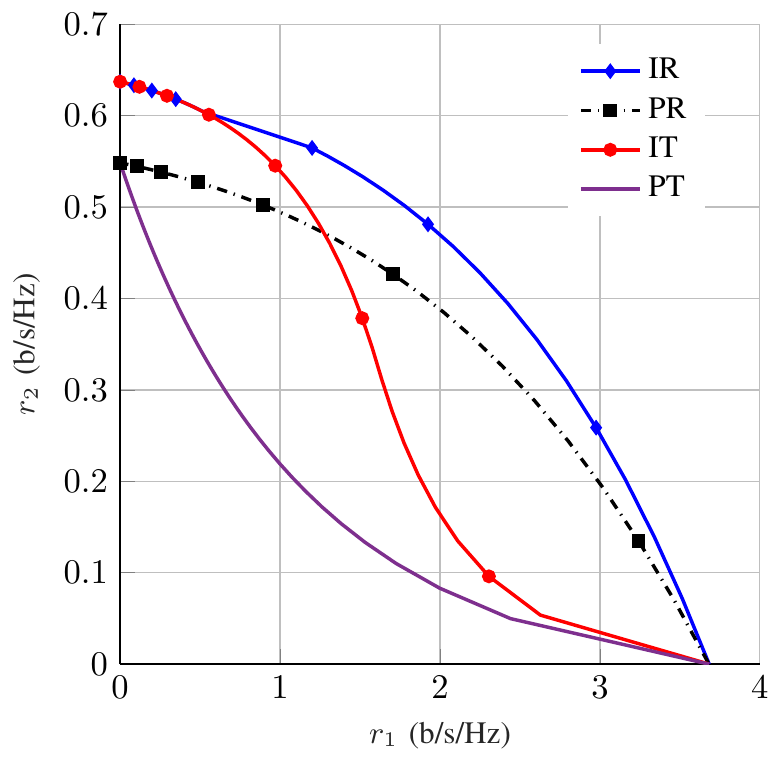}
        \caption{With IQI.}
    \end{subfigure}
\caption{Achievable rate region of a two-user SISO BC with $P=10$ dB and the channel realization $\mathcal{C}_1$.}
	\label{Fig-rr} 
\end{figure}
Fig. \ref{Fig-rr} shows the achievable rate region of a two-user  SISO BC with $P=10$ dB and the channel realization
\begin{align*}
\mathcal{C}_1:\,\,f_1&=-1.3992 + 0.0292i,&f_2&=0.2353 - 0.1238i.
\end{align*}
As can be observed in Fig. \ref{Fig-rr}a, PGS with TIN is very suboptimal, and all other schemes can highly outperform PGS with TIN.  IGS with TIN can enlarge the rate region over the PGS scheme with TIN as well as the TS scheme. Moreover, RS with PGS is the optimal scheme, and  RS with IGS performs the same as RS with PGS when the devices are perfect.
However, it is not the case when there exists IQI, as can be observed in Fig. \ref{Fig-rr}b. 
In the presence of IQI, the noise is improper, and to compensate for it, we should employ improper signaling. As shown in Fig. \ref{Fig-rr}b, IGS with TIN can outperform RS with PGS in some operational points. Additionally, RS with IGS highly outperforms RS with PGS. 
Furthermore, it can be observed in Figs. \ref{Fig-rr}a and \ref{Fig-rr}b that the achievable rate region shrinks when the devices are imperfect.  

\begin{figure}[t!]
            \centering
\includegraphics[width=.25\textwidth]{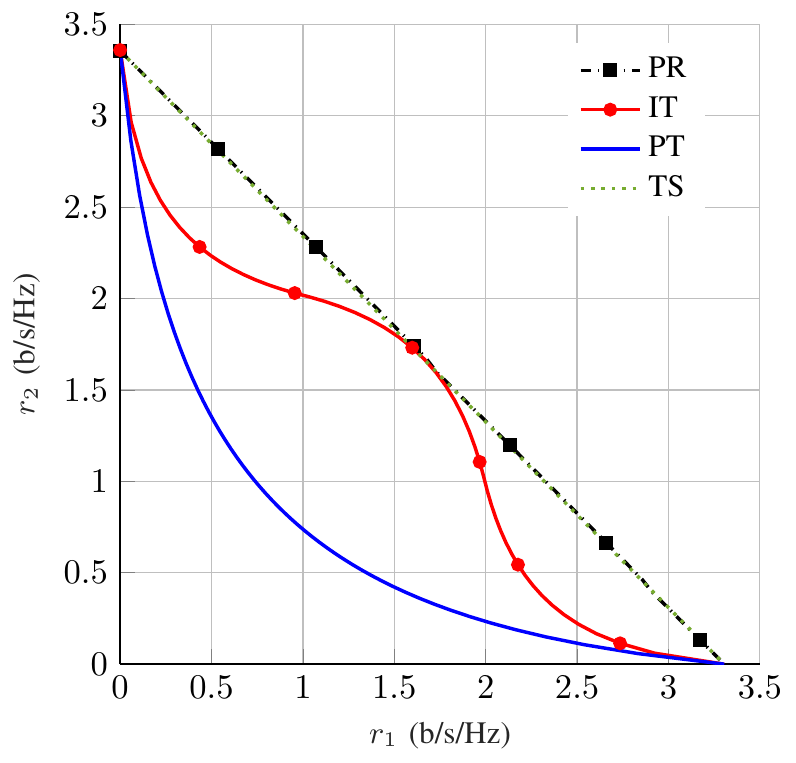}
        \caption{Achievable rate region of a two-user  SISO BC with $P=10$ dB and the channel realization $\mathcal{C}_2$.}
    	\label{Fig-rr-3-s} 
\end{figure}
Fig. \ref{Fig-rr-3-s} shows the achievable rate region of a two-user  SISO BC with $P=10$ dB and the channel realization
\begin{align*}
\mathcal{C}_2:\,\,f_1&=0.3672 + 0.8681i,&f_2&=0.2798 + 0.9214i.
\end{align*}
For this channel realization, the absolute values of the channels are almost equal, and non-orthogonal multiple access schemes cannot provide a significant gain over TDMA and TS. 
As can be observed, RS with PGS is the optimal strategy and attains all the points on the achievable rate region without employing TS. However, PGS and IGS schemes with TIN are highly suboptimal. Indeed, this example shows that the 1-layer RS scheme includes OMA schemes.

\begin{figure}[t!]
    \centering
\includegraphics[width=.3\textwidth]{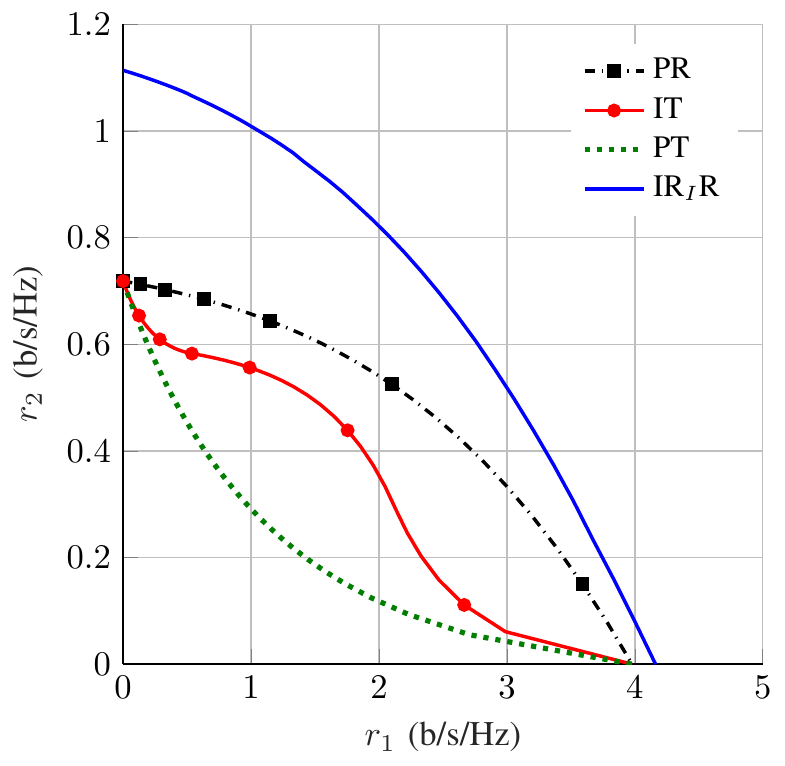}
     \caption{Achievable rate region of a two-user  SISO BC with $P=10$ dB and the channel realization $\mathcal{C}_3$.}
\label{Fig-rr-2}
\end{figure}
Fig. \ref{Fig-rr-2} shows the achievable rate region of a two-user  SISO BC with $P=10$ dB and the channel realization
\begin{align*}
\mathcal{C}_3:\,\,f_1&=0.5909 - 1.0615i,&f_2&=0.2540 - 0.0052i.
\end{align*}
As can be observed, RIS can highly enlarge the achievable rate region. However, we should employ RIS with RS to get the best performance out of RIS. Moreover, we can observe that RIS provides more benefits for the rate of the user with a weaker channel gain, which shown the ability of RIS to significantly improve the coverage.

\subsection{MIMO systems}\label{sec-mc-4-b}
\begin{figure}[t!]
    \centering
    \begin{subfigure}[t]{0.24\textwidth}
        \centering
        \includegraphics[width=.96\textwidth]{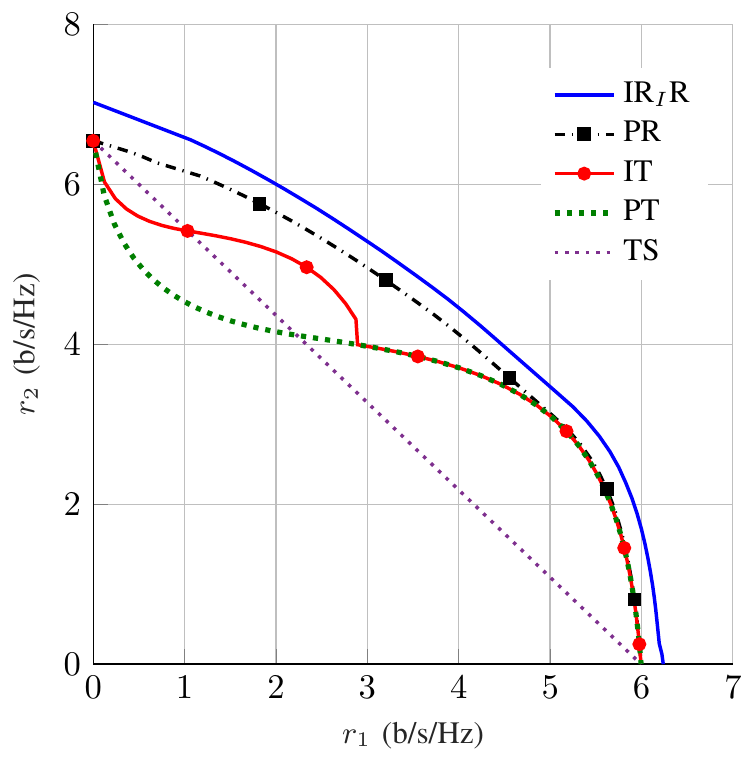}
        \caption{$\mathcal{C}_4$.}
    \end{subfigure}%
    ~ 
    \begin{subfigure}[t]{0.24\textwidth}
        \centering
\includegraphics[width=.96\textwidth]{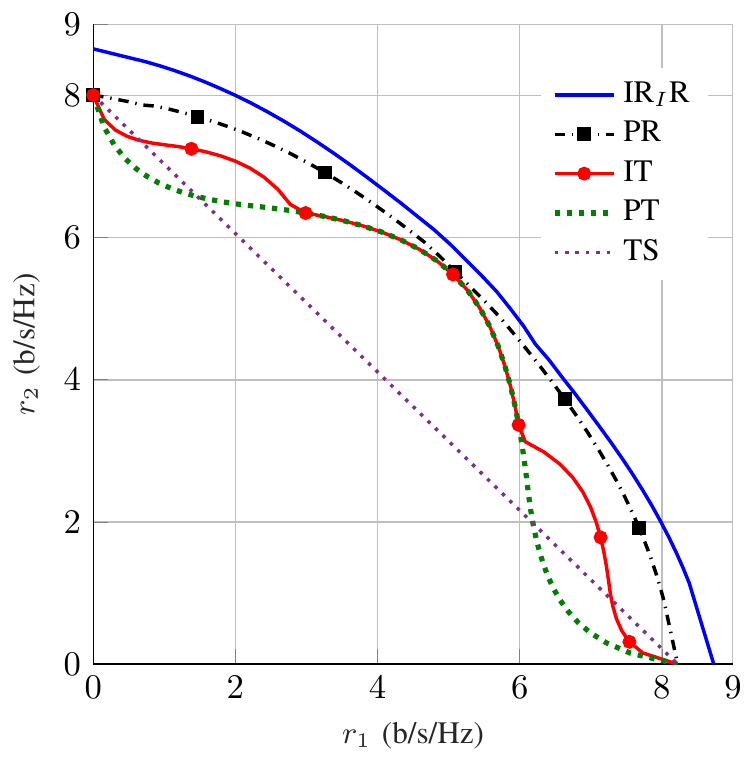}
        \caption{$\mathcal{C}_5$.}
    \end{subfigure}
\caption{Achievable rate region of a two-user $2\times 2$ MIMO BC with $P=10$ dB and different channel realizations.}
	\label{Fig-rr-3} 
\end{figure}
Fig. \ref{Fig-rr-3} shows the achievable rate region of a two-user $2\times 2$ MIMO BC with $P=10$ dB, $\alpha_{RIS}=3$ and the following channel realizations:
\begin{align*}  
\mathcal{C}_4:\,\,&
\mathbf{F}_1=\left[\begin{array}{cc} -1.6952 + 1.7244i&-0.5196 - 0.1194i\\
0.0665 + 0.3475i  & 0.1105 + 0.3237i\end{array}\right],
\\&
\mathbf{F}_2=\left[\begin{array}{cc} 
-0.0233 + 0.6539i  & 0.2841 + 0.8593i\\
  -0.2500 - 1.2059i  & 0.8494 + 0.5047i
\end{array}\right],
\\
\mathcal{C}_5:\,\,&
\mathbf{F}_1=\left[\begin{array}{cc}0.2949 - 0.7399i & -2.1314 + 0.5059i\\
  -1.5491 + 0.3702i & -0.1943 + 0.9528i\end{array}\right],
\\&
\mathbf{F}_2=\left[\begin{array}{cc} 
-0.7849 + 2.4803i &  0.0522 - 0.0681i\\
  -1.5022 + 0.1034i &  0.4433 - 1.0066i
\end{array}\right],
\end{align*}
A  two-user $2\times 2$ MIMO BC can be considered as an underloaded system since the sum of the number of transmit and receive antennas is higher than the number of users. 
As can be observed in Fig. \ref{Fig-rr-3}, IGS with TIN can enlarge the rate region over the PGS with TIN scheme. Furthermore, RS with TIN outperforms the other schemes as it is the optimal scheme in the considered system. We can also observe that RIS can enlarge the rate region by improving the coverage. 
Since this system is underloaded, the benefits of IGS and RS are less than in the two-user SISO BC, which is a highly overloaded system. However, there are still some benefits in the employment of RS and/or IGS.

\section{Summary and conclusion}\label{v}
In this paper, we have characterized the achievable rate region of  
RIS-assisted BCs with RS and IQI.  Our main findings can be summarized as follows:
\begin{itemize}
\item The role of RS is to manage interference. The 1-layer PGS-based RS scheme is optimal in a  two-user BC with perfect devices. This scheme includes OMA, TIN and NOMA. However, when the transceivers suffer from IQI, PGS is unable to compensate for it, and we should employ IGS. Interestingly, IGS with TIN may outperform the 1-layer RS with PGS in some regimes. Thus, in the presence of IQI, the  1-layer IGS-based RS scheme is optimal in a  two-user BC with and/or without RIS.

\item The role of IGS is twofold: to manage interference and to compensate for IQI.  

\item The role of RIS in this system is mainly to improve the coverage, as it cannot completely manage interference in a BC, which is in line with our previous studies \cite{soleymani2022noma, soleymani2022rate}. Indeed, we have to employ advanced interference-management techniques such as RS in highly overloaded systems to use RIS more efficiently.  

\item RS and IGS as interference-management techniques can provide considerable benefits in overloaded systems. However, these benefits decrease (or may even vanish) in underloaded systems.   
\end{itemize}

\section*{Acknowledgment}

The work of I. Santamaria has been partly  supported by the project ADELE PID2019-104958RB-C43, funded by MCIN/AEI/10.13039/501100011033. The work of Eduard Jorswieck was supported in part by the Federal Ministry of Education and Research (BMBF, Germany) as part of the 6G Research and Innovation Cluster 6G-RIC under Grant 16KISK020K.
\bibliographystyle{IEEEtran}
\bibliography{ref2}
\end{document}